\documentclass[a4paper,12pt]{article}

\usepackage{amsmath}
\usepackage{amsfonts}
\usepackage[dvips]{graphicx}
\usepackage[dvips,bookmarksnumbered=true,breaklinks=true]{hyperref}
\usepackage[bindingoffset=0.5cm,textheight=21cm,hdivide={2.8cm,*,2.75cm},vdivide={*,22cm,*}]{geometry}

\usepackage{subfig}

\newcommand{\eqnref}[1]{Eq.~(\ref{#1})}
\newcommand{\figref}[1]{Fig.~\ref{#1}}

\newcommand{\pa}{\partial}

\newcommand{\inv}[1]{\frac{1}{#1}}
\newcommand{\tinv}[1]{\tfrac{1}{#1}}

\newcommand{\intk}{\int\! \frac{d^4k}{\left(2\pi\right)^4}}

\newcommand{\Act}{S}
\newcommand{\ri}{{\rm i}}
\newcommand{\re}{{\rm e}}

\newcommand{\s}{\sigma}

\renewcommand{\l}{\lambda}
\renewcommand{\th}{\theta}
\renewcommand{\k}{\tilde{k}}
\newcommand{\p}{\tilde{p}}
\newcommand{\q}{\tilde{q}}

\newcommand{\uim}{UV/IR mixing}
\newcommand{\nco}{non-commu\-ta\-tive}

\title{\begin{flushright}{\small LYCEN 2008-11}\end{flushright}\vspace{2cm}Quantum Corrections for Translation-Invariant Renormalizable Non-Commutative $\phi^4$ Theory}

\author{Daniel N. Blaschke\footnotemark[1]~,
Fran\c cois Gieres\footnotemark[2]~, Erwin Kronberger\footnotemark[1]~, \\Thomas Reis\footnotemark[1]~, Manfred Schweda\footnotemark[1]~ and Ren\'e I.P. Sedmik\footnotemark[1]}

\date{November 20, 2008}

\graphicspath{{./figures/}}
\begin{document}
\maketitle

\thispagestyle{empty}

\begin{center}
\renewcommand{\thefootnote}{\fnsymbol{footnote}}
\vspace{-0.3cm}\footnotemark[1]Institute for Theoretical Physics,
Vienna University of Technology\\
Wiedner Hauptstrasse 8-10, A-1040 Vienna (Austria)\\[0.3cm]
\footnotemark[2] Universit\'e de Lyon,
Universit\'e Lyon 1 and CNRS/IN2P3,\\
Institut de Physique Nucl\'eaire,
 Bat. P. Dirac,\\
4 rue Enrico Fermi, F - 69622 - Villeurbanne (France)\\[0.5cm]
\ttfamily{E-mail: blaschke@hep.itp.tuwien.ac.at, gieres@ipnl.in2p3.fr,
kronberger@hep.itp.tuwien.ac.at, reis@hep.itp.tuwien.ac.at, mschweda@tph.tuwien.ac.at, sedmik@hep.itp.tuwien.ac.at}
\vspace{0.5cm}
\end{center}
\begin{abstract}
In this paper we elaborate on the translation-invariant renormalizable $\phi^4$ theory in $4$-dimensional  {\nco} space which was recently introduced by the Orsay group. By explicitly performing Feynman graph calculations at one loop and higher orders we illustrate the mechanism which overcomes the {\uim} problem and ultimately leads to a renormalizable model. The obtained results show that the IR divergences are also suppressed in the massless case, which is of importance for the gauge field theoretic generalization of the scalar field model.
\end{abstract}

\pagebreak
\tableofcontents

%%%%%%%%%%%%%%%%%%%%%%%%%%%%%%%%%%%%%%%
\section{Introduction}\label{sec:intro}
%%%%%%%%%%%%%%%%%%%%%%%%%%%%%%%%%%%%%%%
The simplest generalization of the $\phi^4$ theory from ordinary
space to {\nco} space is given by the action~\cite{Minwalla:1999}-\cite{Rivasseau:2007a}
\begin{align}
S_{{\rm naive}} [\phi]& \equiv \int\limits_{\mathbb{R} ^4}
d^4x\left[\inv{2}\left(\pa^\mu\phi\star\pa_\mu\phi
+m^2\phi\star\phi
\right)+\frac{\l}{4!}\,\phi\star\phi\star\phi\star\phi\right]\, .
\label{naive}
\end{align}
Here, $\phi$ denotes a relativistic scalar field in
$4$-dimensional Euclidean space, and the non-commutativity is
implemented by the Weyl-Moyal star
product~\cite{Douglas:2001},
\begin{align}
[ x^{\mu} \stackrel{\star}{,} x^{\nu} ] \equiv  x^{\mu} \star
x^{\nu} -x^{\nu} \star x^{\mu} = \ri \th^{\mu \nu} \, ,
\end{align}
where the parameters $\th^{\mu \nu}= - \th^{\nu \mu}$ are real
constants. In the following, we assume that the deformation
matrix $( \th_{\mu\nu} )$ has the simple block-diagonal form
\begin{align}
( \th_{\mu\nu} )
=\th\left(\begin{array}{cccc}
0&1&0&0\\
-1&0&0&0\\
0&0&0&1\\
0&0&-1&0
\end{array}
\right) \, ,  \qquad {\rm with} \ \; \th \in \mathbb{R} \, .
\end{align}
We note that the parameter $\th$ is necessarily quite small on physical grounds, e.g. see reference~\cite{Nicolini:2008} for a recent discussion of experimental aspects.

By now it is well established that the {\nco} model described by the action (\ref{naive}) is not renormalizable  due to the fact that it suffers from the infamous {\uim} problem that plagues Moyal deformed field theories~\cite{Minwalla:1999,Rivasseau:2007a}. For this reason, we refer to the action (\ref{naive}) as the {\em na\"ive model}~\cite{Rivasseau:2008a}. A procedure to obtain a renormalizable theory consists of adding a properly chosen term to the action (\ref{naive}) in order to overcome the {\uim} problem. In doing so, one clearly changes the initial model, but one obtains a consistent quantum field theory. Recently, three such proposals have been made and each of them has been proved to provide a renormalizable theory (see reference~\cite{Tanasa:2008d} for a short review).

The first proposal, put forward by Grosse and Wulkenhaar~\cite{Grosse:2003}, consists of adding a harmonic oscillator-like potential of the form $\tilde x ^2 \phi
^2$ (where $\tilde x_{\mu} \equiv \th_{\mu \nu} x^{\nu}$). A notable shortcoming of this model is that it breaks translation invariance. We also remark that the corresponding quantum corrections are hard to evaluate due to the occurrence of the
Mehler kernel which is quite involved. The generalization of this model to {\nco} gauge theories is not obvious, though some work in this direction has been done~\cite{Blaschke:2007b}.

Another proposal, due to Grosse and Vignes-Tourneret~\cite{Grosse:2008a}, consists of adding a non-local term of the form $\frac{\mu}{\th^4}\left( \int d^4x \, \phi(x) \right)^2$ to the action (\ref{naive}). This approach yields a minimalist translation-invariant $\phi^4$-theory in {\nco} space.

The third proposal, made by Gurau, Magnen, Rivasseau and Tanasa~\cite{Rivasseau:2008a}, consists of adding a non-local counterterm $\phi\frac{1}{\th^2\square}\phi$ for the quadratic IR divergence of the na\"ive model. This procedure provides a solution for the {\uim} problem while maintaining translation invariance.

In the present work, we focus on the latter model, i.e.
on the action
\begin{align}
\Act[\phi]&\equiv \int\limits_{\mathbb{R} ^4} d^4x \left[\inv{2}
\left(\pa^\mu\phi\star\pa_\mu\phi +m^2\phi\star\phi
-\phi\star\frac{a^2}{\square}\phi\right)
+\frac{\l}{4!}\,\phi\star\phi\star\phi\star\phi\right]\, ,
\label{eq:action}
\end{align}
in Euclidean space. In expression (\ref{eq:action}), the parameter $a$ is assumed to have the form $a=a'/\th$ where $a'$ represents a real dimensionless constant. One of our motivations is that the non-local term has been generalized~\cite{Blaschke:2008a} to $U(1)$ gauge theories whose renormalizability still remains an open problem. Therefore, it should be quite useful to gain deeper insight into the quantum corrections for the model (\ref{eq:action}) since the multiscale analysis~\cite{Rivasseau:2005a} %,Rivasseau:2005b}
which was used to establish its renormalizability cannot be applied to gauge field theories as it breaks gauge invariance. Accordingly we will study perturbative corrections to the propagators and vertices for the theory (\ref{eq:action}), and in particular the vanishing mass limit which is of interest for gauge theories. In doing so, we will explicitly exhibit the improvements for the quantum theory brought about by the non-local term.

Our paper is organized as follows: After introducing the Feynman rules for the model in the next section, we deal with the renormalization procedure at the one-loop level in Sections~\ref{sec:1-loop} and~\ref{sec:remarks}. Section~\ref{sec:higher-loop} is devoted to two and more loops, and exhibits the mechanism that removes the potential IR divergences, which ultimately leads to the renormalizability of the model~\cite{Rivasseau:2008a}. In this respect,  the scaling behaviour of the propagator determined by the action (\ref{eq:action}) plays an essential role. The
appendices describe the techniques for handling the loop calculations and should apply to generic higher loop graphs.

Although our investigations concern the massive theory, we also consider the limit $m\to0$ in several instances and find well defined results. This issue is of importance for the gauge field theoretic generalization of the model (\ref{eq:action}) which was recently proposed~\cite{Blaschke:2008a} and which is to be further discussed elsewhere~\cite{workinprogress}.

\section{Propagator and vertex}\label{sec:model}
%%%%%%%%%%%%%%%%%%%%%%%%%%%%%%%%%%%%%%%%
Concerning the action (\ref{eq:action}) we
recall~\cite{Blaschke:2008a} that the operator $1/\square$ denotes
the Green function associated to the $4$-dimensional Laplacian
$\square\equiv\partial^\mu\partial_\mu=\partial_1^2+\ldots+\partial_4^2$
and that it is given by $-1/k^2$ in momentum space. Accordingly, the  propagator in momentum space reads
\begin{align}
\raisebox{-1.5pt}[1pt][0pt]{\includegraphics[scale=0.8, trim=0 0 0 10,clip=true]{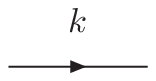}}= \; G(k)=
\inv{k^2+m^2+\frac{a^2}{k^2}}\,.
\label{eq:propagator}
\end{align}
Note that this propagator has a ``damping'' behaviour for
vanishing momentum~\cite{Rivasseau:2008a},
\[
 \lim\limits_{k \to 0}G(k)=0\,,
\]
which allows to avoid potential IR divergences in higher loop graphs (see Section~\ref{sec:higher-loop}). This property is due to the non-local term $\phi\frac{a^2}{\square}\phi$ in the action (\ref{eq:action}) and represents a crucial difference (and improvement) compared to the na\"ive model (\ref{naive}).

In terms of the notation $\k_\mu\equiv\th_{\mu\nu}k^\nu$, the
vertex in momentum space has the following form (see
reference~\cite{Micu:2000}):
%bb=270 675 340 705,scale=0.8 bb=270 620 340 670
\vspace*{4pt}
\begin{align}
\hspace*{-12pt}
\raisebox{-30pt}[24pt][0pt]{\includegraphics[scale=0.8]{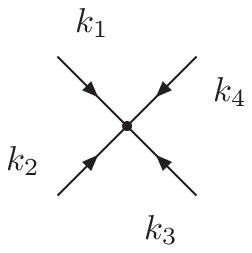}} &=  V(k_1,k_2,k_3,k_4)
\nonumber \\
&=  \frac{\l}{3}(2\pi)^4 \delta^{4}
\left(k_1+k_2+k_3+k_4\right) \Bigg[ \cos \left( \frac{1}{2} \,
k_1\k_2 \right)\cos\left(\frac{1}{2} \,k_3\k_4 \right)
\nonumber\\
& \quad +\cos\left(\frac{1}{2} \, k_1\k_3\right)\cos\left(\frac{1}{2}
\, k_2\k_4\right)+\cos\left(\frac{1}{2} \, k_1\k_4
\right)\cos\left(\frac{1}{2} \, k_2\k_3\right)\Bigg] .
\label{eq:vertex}
\end{align}
%

%%%%%%%%%%%%%%%%%%%%%%%%%%%%%%%%%%%%%%%%%%%%%%%%%%%%%%%%%%%%%%%%%
\section{One-loop corrections for propagator and vertex}\label{sec:1-loop}
%%%%%%%%%%%%%%%%%%%%%%%%%%%%%%%%%%%%%%%%%%%%%%%%%%%%%%%%%%%%%%%%%
To start with, we determine the relevant corrections for the 1PI two-point and four-point functions at one-loop level.

%%%%%%%%%%%%%%%%%%%%%%%%%%%%%%%%%%%%%%%%%%%%%%%%%%%%%%%%%%%%%%%%%
\subsection{Propagator}
%%%%%%%%%%%%%%%%%%%%%%%%%%%%%%%%%%%%%%%%%%%%%%%%%%%%%%%%%%%%%%%%%
\begin{figure}[!ht]
\centering
\includegraphics{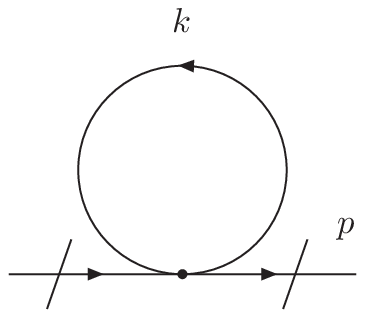}\qquad\includegraphics{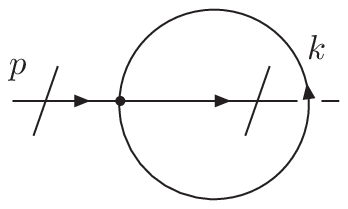}
\caption{Planar and non-planar one-loop corrections for the propagator}
\label{fig:1l-prop}
\end{figure}
The one-loop correction to the propagator, which is described by the Feynman diagrams in \figref{fig:1l-prop}, corresponds to the following integral (including a symmetry factor $1/2$):
\begin{align}
\Pi(p)=- \frac{\l}{6} \int\limits_{\mathbb{R} ^4} \frac{d^4k}{(2\pi)^4} \,
\frac{2+\cos(k\p)}{k^2+m^2+\frac{a^2}{k^2}} \equiv
\Pi^{\text{plan}}+\Pi^{\text{n-pl}}(p)\, . \label{eq:loopint-1}
\end{align}
Here, $\Pi^{\text{plan}}$ and $\Pi^{\text{n-pl}}$ denote the {\em planar} and {\em non-planar parts,} respectively. We note that
\begin{align}
\cos(k\p)= \ & \inv{2}\sum\limits_{\eta=\pm1}
{\rm e} ^{\ri\eta k\p}\, ,
\intertext{and } \inv{k^2+m^2+\frac{a^2}{k^2}}  = \
&\frac{k^2}{\left(k^2+\frac{m^2}{2}\right)^2-M^4}=
\inv{2}\sum\limits_{\zeta=\pm1}\frac{1+\zeta\frac{m^2}{2M^2}}{k^2+\frac{m^2}{2}+\zeta M^2}\, ,
\label{eq:facdec}
\end{align}
where $M^2\equiv\sqrt{\frac{m^4}{4}-a^2}$ (which may be real or purely imaginary depending on the value of $a$).
Taking into account these identities, the {\em non-planar part} can  be evaluated straightforwardly by using Schwinger's exponential parametrization (see Appendix~\ref{app:integrals} for details):
\begin{align}\label{eq:PI_nonpl_result}
&\Pi^{\text{n-pl}}(p) \equiv
-\frac{\l}{24}\intk\sum\limits_{\eta,\zeta=\pm1}
\frac{1+\zeta\frac{m^2}{2M^2}}{k^2+\frac{m^2}{2}+\zeta M^2} \,
\re^{\ri\eta k\p}
\nonumber \\
&= \, -\frac{\l}{48 \pi ^2} \sum\limits_{\zeta= \pm 1}
 \left(
1+\zeta\tfrac{m^2}{2M^2} \right) \, \sqrt{\frac{\frac{m^2}{2}
+\zeta M^2}{\p^{\, 2}}} \; K_{1}\left( \sqrt{ \p^{\, 2}
 \left(\tfrac{m^2}{2}+\zeta M^2\right)} \, \right) \, ,
\end{align}
where $K_1$ is the modified Bessel function. The result is finite for $\p ^{\, 2} \neq 0$, i.e. if $\theta \neq 0$ and $p \neq 0$.

In the following, we will focus on the IR behaviour of the model, i.e. the limit $\p ^{\, 2} \to 0$. For small $z$, the function $\inv{z}K_1(z)$ admits the expansion
\begin{align}\label{bessel-series}
\inv{z}K_1(z)&=\inv{z^2}+\inv{2}\,\ln z+ \frac{1}{2}
\left(\gamma_E - \ln 2 - \frac{1}{2}\right) + \frac{z^2}{16}\left(\ln z+\gamma_E-\ln 2-\frac{5}{4}\right)+\mathcal{O}(z^4) \,,
\end{align}
where $\gamma_E$ denotes the Euler-Mascheroni constant. Thus, for
$\p ^{\, 2} \ll 1$, the expression (\ref{eq:PI_nonpl_result})
behaves like
\begin{align}\label{smallp}
\Pi^{\text{n-pl}}(p)
&=\frac{-\l}{6(4\pi)^2} \Bigg[ \frac{4}{\p^{\, 2}} + m^2 \ln
\left(\p^{\, 2} \sqrt{\tfrac{m^4}{4}-M^4}\right) \nonumber \\
&\quad\hspace{2.3cm}
+\left(M^2+\tfrac{m^4}{4M^2}\right)\ln\sqrt{\frac{\frac{m^2}{2}+M^2}{\frac{m^2}{2}-M^2}}\, \Bigg] +\mathcal{O}(1) \, ,
\end{align}
and thereby involves a quadratic IR divergence (and a subleading logarithmic IR divergence). For $a\to0$ (i.e. $M^2\to\frac{m^2}{2}$) this result reduces to the one which was previously found~\cite{Minwalla:1999,Micu:2000} for $a=0$, i.e. for
the na\"ive model.

The integral defining the {\em planar part} does not contain a phase factor ${\rm e} ^{\ri \eta k\p}$ and is therefore UV divergent. It can be regularized by introducing a cutoff $\Lambda$ and subsequently taking the limit $\p^{\, 2} \to0$, as explained in Appendix~\ref{app:prop}.
The final result can be expanded for large values of $\Lambda$, yielding
\begin{align}\label{eq:PI_planar_result}
\left( \Pi^{\text{plan}} \right)_{{\rm regul.}} (\Lambda )
&= \frac{-\l}{3(4\pi)^2} \Bigg[ 4\Lambda^2+m^2\ln
\left(\tinv{\Lambda^2}\sqrt{\tfrac{m^4}{4}-M^4}\, \right)
\nonumber\\
&\quad\hspace{2.3cm} +\left(M^2+\tfrac{m^4}{4M^2}\right)
\ln\sqrt{\frac{\frac{m^2}{2}+M^2}{\frac{m^2}{2}-M^2}} \, \Bigg]
+\mathcal{O}(1).
\end{align}

%%%%%%%%%%%%%%%%%%%%%%%%%%%%%%%%%%%%%%%%%%%%%%%%%%%%%%%%%%%%%%%%%
\subsection{Vertex}
%%%%%%%%%%%%%%%%%%%%%%%%%%%%%%%%%%%%%%%%%%%%%%%%%%%%%%%%%%%%%%%%%
The basic one-loop correction to the vertex is given by the three connected graphs that can be constructed with four external legs~\cite{Maggiore:2005,Micu:2000,Rim:2003}:
\begin{align}
V_{\text{1-loop}} (p_1,p_2,p_3,p_4) &=
\tinv{3}\Bigg[\tinv{2}\raisebox{-28pt}[24pt][0pt]{\includegraphics[scale=0.8]{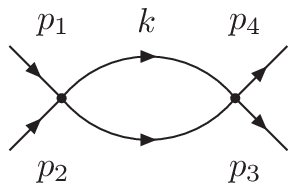}}
{+} \raisebox{-23pt}[24pt][0pt]{\includegraphics[scale=0.8, trim=20 0 25 0, clip=true]{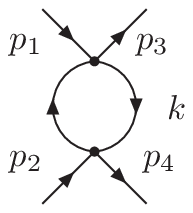}} + \raisebox{-23pt}[24pt][0pt]{\includegraphics[scale=0.8, trim=22 0 25 0, clip=true]{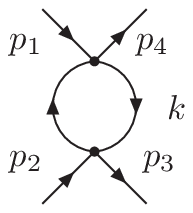}}\Bigg].
\label{eq:vertex_correction_complete}
\end{align}
This expression can be evaluated by proceeding along the lines of
reference~\cite{Micu:2000}: by applying the Feynman rules
(\ref{eq:propagator}) and (\ref{eq:vertex}), and by taking advantage
of the identity (\ref{eq:facdec}) we find that
(\ref{eq:vertex_correction_complete}) reads
\begin{align}
&  \frac{\lambda^2}{27}
\sum\limits_{\zeta,\chi=\pm1}  \intk
\frac{\left(1+\zeta\frac{m^2}{2M^2}\right)
\left(1+\chi\frac{m^2}{2M^2}\right)}{k^2+\frac{m^2}{2}+\zeta M^2}
\Bigg[ \left(1+\tinv{4}\sum\limits_{i=2}^4\re^{\ri k(\p_1+\p_i)}+\tinv{2}\sum\limits_{i=1}^4\re^{\ri k\p_i}\right)
\nonumber\\
&\hspace*{4.0ex} \times \left(\tinv{(p_1+p_2-k)^2+\frac{m^2}{2}+\chi M^2}+\tinv{(p_1+p_3-k)^2+\frac{m^2}{2}+\chi M^2}+\tinv{(p_1+p_4-k)^2+\frac{m^2}{2}+\chi M^2}\right)
\nonumber\\
&\hspace*{2.5ex}
+\tfrac{3}{4}\left(\tfrac{\displaystyle{\re}^{\ri k(\p_1+\p_2)}}{(p_1+p_2-k)^2+\frac{m^2}{2}+\chi M^2}+\tfrac{\displaystyle{\re}^{\ri k(\p_1+\p_3)}}{(p_1+p_3-k)^2+\frac{m^2}{2}+\chi M^2}+\tfrac{\displaystyle{\re}^{\ri k(\p_1+\p_4)}}{(p_1+p_4-k)^2+\frac{m^2}{2}+\chi M^2}\right)\Bigg].
\label{eq:vertex_correction_complete_eq}
\end{align}
Thus, we again have an expression involving planar and non-planar parts (the latter involving a phase factor of the form $\re^{\ri k\q}$). The generic integral for the {\em non-planar part} is given by
\begin{align}
& I(p,q) \equiv  \sum\limits_{\zeta,\chi=\pm1} \intk
\frac{\left(1+\zeta\frac{m^2}{2M^2}\right)\left(1+\chi\frac{m^2}{2M^2}\right)\re^{\ri k(\p+\q)}}{\left(k^2+\frac{m^2}{2}+\zeta M^2\right)
\left[\left(p-k\right)^2+\frac{m^2}{2}+\chi M^2\right]}\nonumber\\
&= \sum\limits_{\zeta,\chi} \left(1+\zeta\tfrac{m^2}{2M^2}\right)
\left(1+\chi\tfrac{m^2}{2M^2}\right) \int\limits_{0}^{1}d\xi \,
\frac{{\rm e}^{\ri (1-\xi)p\q}}{8\pi^2} \nonumber\\
&\quad\hspace{1.7cm}\times K_0\left(\sqrt{(\p+\q)^2\left[\xi(1-\xi)p^2
+\frac{m^2}{2}+\left(\chi+\xi(\zeta-\chi)\right)M^2\right]} \,
\right).
\label{ver-int1}
\end{align}
Here, $p$ denotes the total incoming momentum, and $q$ represents one of the variables $p_i$ (see Appendix~\ref{app:integrals} for calculational details).
For small arguments the modified Bessel function $K_0$ can be expanded according
to
\begin{align}
K_0(z)&= -\ln z  + \ln 2 -\gamma_E +\mathcal{O}\left(z^2\right) \,,
\end{align}
from which we can derive the following estimation for small external momenta $p$ and $q$:
\begin{align}\label{vertex-result-1loopeq:vertex_correction_complete}
|I(p,q)|&\leq \frac{1}{(2\pi)^2}\Bigg|\ln\left(\frac{(\p+\q)^2}{4}\sqrt{\tfrac{m^4}{4}-M^4}\right)+2\gamma_E-\inv{2}\left(1-\tfrac{m^4}{4M^4}\right)\nonumber\\
&\qquad \qquad\qquad \qquad \qquad \qquad  \qquad
+\left(1+\tfrac{m^4}{4M^4}\right)\tfrac{m^2}{4M^2}
\ln\sqrt{\frac{\frac{m^2}{2}+M^2}{\frac{m^2}{2}-M^2}}\Bigg|.
\end{align}
The {\em planar part} of expression (\ref{eq:vertex_correction_complete}) can again be evaluated by introducing a cut-off $\Lambda$ (as was discussed for the propagator
in Appendix A.1): the final result directly follows from (\ref{ver-int1}) and (\ref{vertex-result-1loopeq:vertex_correction_complete}) by replacing  $(\p+\q)^2$  with $1 / \Lambda ^2$.

%%%%%%%%%%%%%%%%%%%%%%%%%%%%%%%%%%%%%%%%%%%%%%%
\section{One-loop renormalization}\label{sec:remarks}
%%%%%%%%%%%%%%%%%%%%%%%%%%%%%%%%%%%%%%%%%%%%%%%
According to the standard renormalization procedure, the {\em dressed propagator at one-loop level} is given by
\begin{align}\label{eq:dressed1}
\raisebox{-13pt}[19pt][0pt]{\includegraphics[scale=0.8]{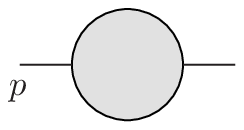}}
\equiv \Delta'(p)&= \inv{A} + \inv{A}\,
\Sigma(\Lambda,p)\, \inv{A} \, ,
\end{align}
where
\begin{align*}
& A \equiv p^2+m^2+\frac{a^2}{p^2} \, ,
\\
& \Sigma(\Lambda,p)\equiv \left( \Pi^{\text{plan}}
\right)_{{\rm regul.}} (\Lambda) +\Pi^{\text{n-pl}}(p)\,.
\end{align*}
Since $A\neq 0$, we can apply the formula
\begin{align*}
\inv{A+B}=\inv{A}-\inv{A}\, B \, \inv{A+B} = \inv{A}-\inv{A}\, B
\, \inv{A} + {\cal O} (B^2) \, ,
\end{align*}
which allows us to rewrite expression \eqref{eq:dressed1} to order
$\Sigma$ (i.e. to order $\lambda$) as
\begin{align}
\Delta'(p)&=\inv{p^2+m^2+\frac{a^2}{p^2}-\Sigma(\Lambda,p)} \, .
\end{align}
The contribution $\Pi^{\text{n-pl}}(p)$ to $\Sigma(\Lambda,p)$ is finite except for vanishing external momentum $p$. The expansion for small values of $\p^{\, 2}$, as given in \eqnref{smallp}, reveals a quadratic and a logarithmic IR divergence\footnote{In this respect, we should emphasize that these IR divergences are fundamentally different from the ones encountered in quantum field theories on commutative space since they are tied to the UV divergences and only appear in non-planar diagrams which are not present in usual QFT~\cite{Douglas:2001}. Thus, these divergences cannot be regularized by introducing an infrared regulator (like an additional mass parameter).} at $\p^{\, 2}=0$. The quadratically divergent term obviously has the same structure as the term $\frac{a^2}{p^2}$ appearing in the bare propagator (\ref{eq:propagator}). Henceforth, we will absorb it by a finite renormalization of  the parameter $a^2$.

From the expansions (\ref{smallp}) and (\ref{eq:PI_planar_result}) it follows that, to order $\lambda$, we have
\begin{align}
\Delta'(p)&= \frac{Z}{p^2+m^2_r+\frac{a^2_r}{p^2}+f(p^2)}\,,
\end{align}
where
\begin{align}\label{renorm-para}
& Z \equiv 1 + \lambda \alpha \theta ^2
\, , \qquad \qquad (\alpha \in \mathbb{R})
\,,\nonumber\\
& m^2_r \equiv
m^2+\frac{\lambda}{3(4\pi)^2}
\left[4\Lambda^2+m^2\ln\left(\tinv{\Lambda^2}\sqrt{\tfrac{m^4}{4}-M^4}\right)\right]
+\mbox{regular for $\Lambda \to \infty$}
\,,\nonumber\\
& a^2_r\equiv a^2+  \l \left[ \frac{2}{3(4\pi\th)^2} + \alpha a^2 \theta ^2 \right]
\,,\nonumber\\
& f(p^2) \equiv  \frac{\lambda}{6 (4 \pi )^2}
\left[ m^2 \, \ln\left(\theta^2 p^2 \right) + {\cal O}((\theta p)^4)  \right] \, .
\end{align}
The quantities $m_r$ and $a_r$ represent the renormalized mass and $a$-parameter to one-loop order, and the function $f(p^2)$ is analytic for $\theta \neq 0$ and $p^2 >0$. The expression $Z$ amounts to a finite wave function renormalization\footnote{For the $\phi ^4$-theory on commutative space, there is no wave function renormalization at one-loop order, but this is a peculiarity of this theory~\cite{Maggiore:2005}.}. The logarithmic singularity of $f(p^2)$ for vanishing external momentum $p$ represents a mild divergence which is unproblematic for the amplitudes (see also the multiscale analysis~\cite{Rivasseau:2008a}). The constant $\alpha$ appearing in $Z$ and in
\begin{align}
\label{num} a^{\prime2}_r \equiv \theta^2 a_r^2 = a^{\prime2}+ \l \left[ \frac{2}{3(4\pi)^2} + \alpha  \theta ^2 a^{\prime2}\right]
\end{align}
is determined by the numerical factor that occurs in the expansion of $\Sigma(\Lambda,p)$ at order $\p^{\, 2}$ (see equations (\ref{bessel-series}) and (\ref{smallp})). We have
\begin{align}
\alpha \theta^2 =\frac{2}{3(16\pi)^2} \left(\ln2+\frac{5}{4}-\gamma_E\right) \left(\theta^2 m^4-a'^2\right)\,,
\end{align}
which is positive for $\theta^2m^4>a'^2$. However, even in the case where $\alpha <0 $, the one-loop renormalized parameter $a^{\prime2}_r$ is positive provided
\begin{align}
a^{\prime2}&<\inv{2}\left(\frac{B}{\l A}+\th^2m^4\right) +\sqrt{\inv{4}\left(\frac{B}{\l A}+\th^2m^4\right)^2+B} \,,\nonumber\\
\text{where} \qquad A&\equiv \frac{2}{3(4\pi)^2}
\, , \quad B\equiv \frac{16}{\ln2+\frac{5}{4}-\gamma_E} >0 \, .
\end{align}
Since $\th$ is necessarily quite small on physical grounds, the dominating factor in the previous inequality is $1/\l$. Hence, even for $m=0$, the parameter $a^{\prime2}_r$ is positive for small values of the coupling constant $\l$ (more precisely for $a'^2<10^3/\l$).

The renormalized coupling constant $\lambda_r$ at one-loop order is obtained by considering the planar part of (\ref{eq:vertex_correction_complete}). One finds an expression which is similar to the one in the commutative theory.
The non-planar part of (\ref{eq:vertex_correction_complete}) again involves a logarithmic singularity (see Equation (\ref{vertex-result-1loopeq:vertex_correction_complete})).

%%%%%%%%%%%%%%%%%%%%%%%%%%%%%%%%%%%%%%%%%%%%%%%%
\section{Two and higher loops}\label{sec:higher-loop}
\begin{figure}[ht]
\centering
\subfloat[with 1 insertion]{\raisebox{37.2pt}{\includegraphics{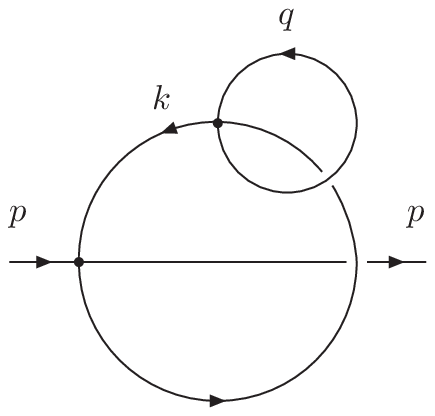}}\label{fig:snowman}}\quad\qquad
\subfloat[with 3 insertions]{\includegraphics{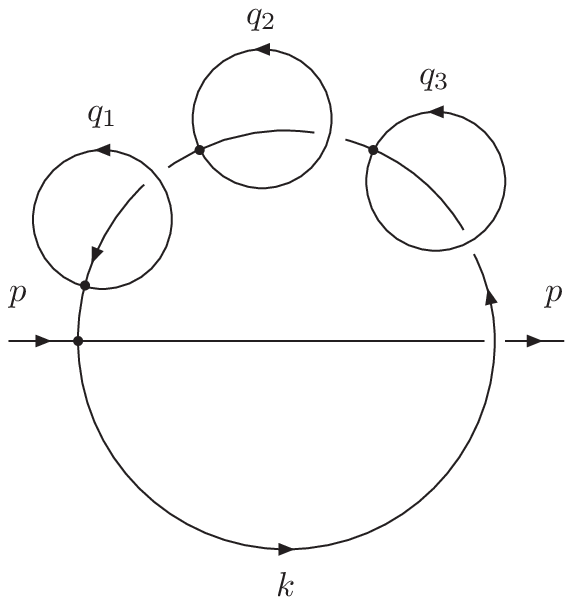}\label{fig:n-insertions}}
\caption{Non-planar 2-loop and 4-loop graphs.} \label{fig:ins}
\end{figure}

%%%%%%%%%%%%%%%%%%%%%%%%%%%%%%%%%%%%%%%%%%%%%%%%
In order to exhibit how the non-local term $\phi\frac{1}{\th^2\square}\phi$  improves the IR behaviour of the na\"ive model at the higher loop level, we consider a non-planar tadpole graph with non-planar insertions (see \figref{fig:snowman} for one insertion and \figref{fig:n-insertions} for several insertions).
Since we are only concerned about the IR divergences, we limit
ourselves to the first (i.e. most singular) term in the expansion
(\ref{smallp}) of $\Pi^{\text{n-pl}}$,  hence~\cite{Minwalla:1999}
we consider  the approximation $\Pi^{\text{n-pl}}(k) \propto
1/\k^2$. Within this approximation, a graph with $n$ non-planar
insertions is described by the expression

\begin{align}\label{n-loop-int}
\Pi^{n \text{ npl-ins.}}(p)&\equiv\l^2\sum\limits_{\eta=\pm1}\intk
\, \frac{\re^{\ri\eta k\p}}{\left(\k^2\right)^n\left[k^2+m^2+\frac{a^2}{k^2}\right]^{n+1}}\,.
\end{align}
For the na\"ive model (where  $a=0$), the integral (\ref{n-loop-int}) involves an IR divergence for $n\geq2$, because the integrand behaves like $(k^2)^{-n}$ for $k^2\to 0$. In contrast, for the model under consideration (where  $a\neq 0$), the integrand behaves like
\begin{align}
\inv{\left(\k^2\right)^n\left[\frac{a^2}{k^2}\right]^{n+1}}=\frac{\k^2}{\left(a'^2\right)^{n+1}}\,.
\end{align}
Thus, the propagator (\ref{eq:propagator}) ``damps'' the IR-dangerous insertions and therefore cures potential IR problems in the integral (\ref{n-loop-int}). This is a nice demonstration of the mechanism leading to the renormalizability of the present model. In this respect we recall that its renormalizability has been proved quite generally in reference~\cite{Rivasseau:2008a} using multiscale analysis.

In the following we will present a more detailed mathematical analysis of the IR behaviour of the graph with $n$ non-planar insertions. In fact, the integral (\ref{n-loop-int}) can be evaluated by the same techniques as those applied  in the previous section (see Appendices~\ref{app:2l-integrals} and~\ref{app:nl-integrals} for details).
For $n=1$, one finds that
\begin{align}\label{2loopresult}
\Pi^{1 \text{ npl-ins.}}(p)&=\frac{-\l^2}{  16 \pi^2\th^2 M^6}
\Bigg\{ m^2 \left[ \sqrt{\frac{m_+^2}{\p^{\,2}} } \, K_1 \left(
\sqrt{ m_+^2 \p^{\,2} } \right) - \sqrt{\frac{m_-^2}{\p^{\,2}} }
\, K_1 \left( \sqrt{ m_-^2 \p^{\,2} } \right)
\right] \nonumber \\
&\quad\hspace{1.8cm} +M^2 \left[ m_+^2 \, K_0 \left(\sqrt{ m_+^2
\p^{\,2} } \right) + m_-^2 \, K_0 \left(\sqrt{ m_-^2 \p^{\,2} }
\right) \right] \Bigg\} \,,
\end{align}
where $m_{\pm}^2 \equiv \frac{m^2}{2} \pm M^2$. For generic $n$,
we get
 $\Pi ^{n \text{ npl-ins.} }(p) = \l ^2 \sum\limits_{\eta = \pm
1} J_n (p)$ with $J_n (p)$ given by the integral (\ref{jnp}). By
expanding the expressions for $n=1$ and $n=2$ for small external
momentum $\p^{\,2}$, one obtains
\begin{align}\label{2-l-result}
\Pi^{1 \text{ npl-ins.}}(p)&=\frac{\l^2}{  16 \,
\pi^2\th^2M^6} \Bigg[\left(M^4-\tfrac{m^4}{4}\right)
\ln\sqrt{\frac{m_+^2}{m_-^2}}
+M^2\tfrac{m^2}{2}\Bigg]+\mathcal{O}( \p^{\, 2}  )\,,
\\
\Pi^{2 \text{ npl-ins.}} (p) &= \frac{\l^2}{256 \, \pi^2 \th^4}
\left[\frac{3m^4-4M^4}{M^{10}} \ln\sqrt{\frac{m_+^2}{m_-^2}}
-6\frac{m^2}{M^8}\right]+\mathcal{O}(\p^{\, 2})\,. \nonumber
\end{align}
We note that the latter result diverges in the limit $a\to0$. This fact again illustrates how the propagator (\ref{eq:propagator}) regularizes graphs which diverge in the na\"ive model.

In the limit $m\to0$ (i.e. for a massless field), the expressions
(\ref{2-l-result}) reduce to finite quantities:
\begin{align}
\Pi^{1 \text{ npl-ins.}}(p)\Bigg|_{m=0}&=\frac{\l^2}{ 32 \, \pi \,
|\th a'| }+\mathcal{O}(\p^{\,2})\,,
\nonumber
\\
\Pi^{2 \text{ npl-ins.}}(p)\Big|_{m=0}&= \frac{ \l^2}{128 \, \pi |\th a'^3|} +\mathcal{O}(\p^{\, 2})\,.
\label{J2}
\end{align}
Henceforth, in contrast to the na\"ive model (e.g. see reference~\cite{Micu:2000}), the higher-loop graphs in \figref{fig:n-insertions} do \emph{not} diverge for $m\to 0$. In other words, the IR divergent insertion $1/\k^2$ does not cause any harm in these higher-loop graphs, even for a massless field.
This is an important feature for the gauge field theoretic generalization of the model~(\ref{eq:action}) which was introduced in reference~\cite{Blaschke:2008a}, and which is to be addressed elsewhere~\cite{workinprogress}.

%%%%%%%%%%%%%%%%%%%%%%%%%%%%%%%%%%%%%%%%%%%%%%%%%
\section{Concluding remarks}\label{concl}
%%%%%%%%%%%%%%%%%%%%%%%%%%%%%%%%%%%%%%%%%%%%%%%%%%
Concerning the quantum corrections for the model under consideration, we note  the quite recent work~\cite{Tanasa:2008a,Tanasa:2008b,Tanasa:2008c} which is complementary to our study: the second work is devoted to the calculation of the one-loop beta function and to the parametric representation of Feynman amplitudes (according to the topology of the considered graphs). The approach of references~\cite{Tanasa:2008a,Tanasa:2008b} is based on the assumption that $4a^2<m^4$ whereas our analysis does not require any restrictions on the parameters of the theory (as was mentioned in Section~\ref{sec:1-loop}).

While some of the obtained expressions allow for a smooth limit $a\to 0$ towards the corresponding results of the na\"ive (non-renormalizable) theory, this is --- as
expected --- not the case for the higher loop corrections.

As is apparent from the calculations outlined in the appendices, the Schwinger parame\-trization is quite convenient for determining the quantum corrections for the scalar field model. Since the gauge field theoretic generalization of this model extends the propagator for a (massless) scalar field, the same techniques should allow us to tackle the problem of IR divergences and discuss the issue of renormalizability for translation-invariant gauge field theories on {\nco} space.

\section*{Acknowledgments}
The work of D.~N.~Blaschke, E.~Kronberger, T.~Reis and R.~I.~P.~Sedmik was supported by the ``Fonds zur F\"orderung der Wissenschaftlichen Forschung'' (FWF) under contracts P20507-N1 and P19513-N16. F.~Gieres also acknowledges the financial support of this program for his visit to the Vienna University of Technology, during which part of the present work has been carried out. He furthermore acknowledges partial support by the ``Agence Nationale de la Recherche'' under contract ANR-07-CEXC-010.

The authors would like to thank F. Delduc, H. Grosse and M. Wohlgenannt for enlightening discussions.

%%%%%%%%%%%%%%%%%%%%%%%%%%%%%%%%%%%%%%%%%%%%%%%%%
\appendix
\section{1-loop integrals}\label{app:integrals}
\subsection{Correction to the propagator}\label{app:prop}
%%%%%%%%%%%%%%%%%%%%%%%%%%%%%%%%%%%%%%%%%%%%%%%%%%
By virtue of equations (\ref{eq:loopint-1})-(\ref{eq:facdec}), the
non-planar part of the one-loop correction to the propagator reads $\Pi^{\text{n-pl}}(p) = -\frac{\l}{24} \, \sum\limits_{\eta=
\pm 1} I(p)$ with
\begin{align}\label{defI}
I(p) & \equiv \sum\limits_{\zeta=\pm 1}\left(1+\zeta\tfrac{m^2}{2M^2}\right)\intk
\frac{\re^{\ri\eta k\p}}{k^2+\frac{m^2}{2}+\zeta M^2}
\, .
\end{align}
For $m>0$ and $a\neq0$, the combination $\frac{m^2}{2} +\zeta M^2$
has a positive real part so that we can use the Schwinger
parametrization:
\begin{align}
\inv{k^2+\frac{m^2}{2}+\zeta M^2}
&=\int\limits_0^{\infty}d\alpha
\, \re^{-\alpha (k^2+\frac{m^2}{2}+\zeta M^2)}
\,.
\end{align}
The integral over $k$ can be carried out after completing the
square in the exponent:
\begin{align}\label{app-integral1}
I(p)
&=\sum\limits_{\zeta}
\left(1+\zeta\tfrac{m^2}{2M^2}\right)\intk\int
\limits_{0}^{\infty}d\alpha
\exp\left[-\alpha\left(k^2-\frac{\ri\eta k \p}{\alpha}\right)
-\alpha\left(\frac{m^2}{2}+\zeta M^2\right)\right]
\nonumber
\\
&=\sum\limits_{\zeta}
\frac{1+\zeta\frac{m^2}{2M^2}}{(4\pi)^2} \int \limits_{0}^{\infty}
\frac{d\alpha}{\alpha^2} \, \exp\left[-\frac{\p^{\,
2}}{4\alpha}-\alpha\left(\frac{m^2}{2} +\zeta M^2\right)\right] \,
.
\end{align}
Although it is not necessary here, we introduce a cutoff $\Lambda$
into $I(p)$ by making the replacement
$\p^{\,2}\to\p^{\,2}+\inv{\Lambda^2}$. In fact~\cite{Micu:2000},
this will allow us to evaluate the planar part $\Pi^{\text{pl}}$
below.  The integral (\ref{app-integral1}) can be looked up
\cite{Gradshteyn:2007} and the result depends upon the modified
Bessel function $K_{-1}=K_1$:
\begin{align}\label{bess}
I_{\text{regul.}}(p,\Lambda) =  \,  \sum\limits_{\zeta}
\frac{1+\zeta\frac{m^2}{2M^2}}{(2\pi )^2}\, \sqrt{\frac{\frac{m^2}{2}
+\zeta M^2}{\p^{\, 2}+\inv{\Lambda^2}}} \; K_{1}\left(
\sqrt{\left(\p^{\, 2}+\tinv{\Lambda^2}\right)
 \left(\tfrac{m^2}{2}+\zeta M^2\right)} \, \right).
\end{align}
Taking the limit $\Lambda\to\infty$ leads to the result
(\ref{eq:PI_nonpl_result}) for  the non-planar part
$\Pi^{\text{n-pl}}(p)$.   The result (\ref{eq:PI_planar_result})
for the regularized planar part
$\left(\Pi^{\text{plan}}\right)_{\text{regul.}}(\Lambda)$ also
follows from the expression (\ref{bess}) by taking the limit
$\p^{\, 2} \to0$.

%%%%%%%%%%%%%%%%%%%%%%%%%%%%%%%%%%%%
\subsection{Vertex correction}
%%%%%%%%%%%%%%%%%%%%%%%%%%%%%%%%%%%%
The non-planar integrals in the vertex correction (\ref{eq:vertex_correction_complete_eq}) have the following form:
\begin{align}
I(p,q) \equiv
& \sum\limits_{\zeta,\chi}
\intk
\frac{\left(1+\zeta\frac{m^2}{2M^2}\right)\left(1+\chi\frac{m^2}{2M^2}\right)
\re^{\ri k(\p+\q)}}{\left( k^2+\frac{m^2}{2}+\zeta M^2 \right)
\left(\left(p-k\right)^2+\frac{m^2}{2}+\chi M^2 \right)}
\nonumber
\\
= & \sum\limits_{\zeta,\chi}
\left(1+\zeta\tfrac{m^2}{2M^2}\right)\left(1+\chi\tfrac{m^2}{2M^2}\right)
 \intk\int\limits_0^{\infty} d\alpha \int\limits_0^{\infty}
d\beta \exp\Big[-(\alpha+\beta)\frac{m^2}{2} \nonumber \\
& \quad \hspace{0.8cm} -(\alpha+\beta)\left(k^2-\frac{\ri k(\p+\q) +2\beta
kp}{\alpha+\beta}\right)  -\alpha\zeta
M^2-\beta\left(p^2+\chi M^2\right) \Big] .
\end{align}
After carrying out the integration over $k$, and performing the
change of variables $(\alpha , \beta ) \to ( \l , \xi)$ with
$\alpha = \l \xi $ and $\beta = \l (1 -\xi)$ (where $\l \in [0,
\infty[$ and $\xi \in [0, 1 ]$), one obtains an integral over
$\lambda$,
\begin{align}\label{eq:1l-vertex-result_1}
I(p,q) &= \inv{(4\pi)^2} \sum\limits_{\zeta,\chi}
\left(1+\zeta\tfrac{m^2}{2M^2}\right)\left(1+\chi\tfrac{m^2}{2M^2}\right)
 \int\limits_{0}^{1}d\xi\int\limits_{0}^{\infty}
\frac{d\l}{\l} \,
\\
& \quad \times  \exp\left\{-\tfrac{(\p+\q)^2}{4\l}+\ri (1-\xi)p\q
\nonumber
 -\l\left[\xi(1-\xi)p^2
+\tfrac{m^2}{2}+\left(\chi+\xi(\zeta-\chi)\right)M^2\right]\right\},
\nonumber
\end{align}
which can be expressed~\cite{Gradshteyn:2007} in terms of the modified Bessel function $K_0$,  see \eqnref{ver-int1}.

%%%%%%%%%%%%%%%%%%%%%%%%%%%%%%%%%%%%%%%%%%%%%%%%%%
\section{2-loop integrals}\label{app:2l-integrals}
%%%%%%%%%%%%%%%%%%%%%%%%%%%%%%%%%%%%%%%%%%%%%%%%%%
We use the decomposition (\ref{eq:facdec}) and the Schwinger
parametrization  to evaluate the integral (\ref{n-loop-int}) for
$n=1$. Hence $\Pi^{1 \text{ npl-ins.}}(p) = \l ^2 \,
\sum\limits_{\eta= \pm 1} J_1(p)$ with
\begin{align}
J_1 (p)
&\equiv\inv{4}\sum\limits_{\zeta,\chi}\intk\frac{\re^{\ri\eta k\p}}{\k^2}\,\frac{1+\zeta\frac{m^2}{2M^2}}{k^2+\frac{m^2}{2}+\zeta M^2}\,\frac{1+\chi\frac{m^2}{2M^2}}{k^2+\frac{m^2}{2}+\chi M^2}\nonumber\\
&=\inv{4\th^2}\sum\limits_{\zeta,\chi}\left(1+\zeta\tfrac{m^2}{2M^2}\right)\left(1+\chi\tfrac{m^2}{2M^2}\right)\intk\int\limits_0^{\infty}d\alpha \int\limits_0^{\infty}d\beta \int\limits_0^{\infty}d\gamma \nonumber\\
&\quad\hspace{1.7cm}\times\exp\Big[-(\alpha+\beta+\gamma)k^2-(\alpha+\beta)\tfrac{m^2}{2}-(\alpha\zeta +\beta\chi)M^2+\ri\eta k\p\Big]\,.
\end{align}
After carrying out the integration over $k$ and after the  change
of variables
\begin{align}\label{eq:lambda_xi_sigma_param}
(\alpha,\beta,\gamma)&\to(\lambda,\xi,\s)\quad\nonumber\\
\text{with}\quad\alpha&=\l \xi\s \,,\nonumber\\
\beta&=\l (1-\xi)\s\,,\nonumber\\
\gamma&=\l (1-\s) \,,\nonumber\\
\text{and}\quad\l &\in [0, \infty[\,,\ \xi \in [0, 1 ]\,,\ \s \in [0, 1 ]\,,
\end{align}
one obtains
\begin{align}\label{Jp-2l}
J_1 (p)
&=\sum\limits_{\zeta,\chi}\frac{\left(1+\zeta\tfrac{m^2}{2M^2}\right)\left(1+\chi\tfrac{m^2}{2M^2}\right)}{4\th^2(4\pi)^2}\int\limits_0^{\infty}d\l\int\limits_0^1d\xi \int\limits_0^1d\s\s\nonumber\\
&\quad\hspace{3.3cm}\times\exp\Big[-\frac{\p^{\,2}}{4\l}
-\l\s\left(\tfrac{m^2}{2}+\xi\zeta M^2+(1-\xi)\chi M^2\right)\Big]
\nonumber\\
&=\inv{\th^2(4\pi)^2}\int\limits_0^{\infty}d\l\int\limits_0^1d\xi
\int\limits_0^1d\s\s
\left[\cosh\left(\l\s \xi M^2 \right)
-\frac{m^2}{2M^2}\sinh\left(\l\s \xi M^2\right)\right]\nonumber\\
&\quad\hspace{1.3cm}\times \left[\cosh\left(\l\s (1-\xi)
M^2\right) -\frac{m^2}{2M^2}\sinh\left(\l\s (1-\xi)M^2
\right)\right]\re^{-\frac{\p^{\,2}}{4\l}-\l\s\tfrac{m^2}{2}}\,.
\end{align}
After integrating out $\xi$ and $\s$, one is left with
sums of integrals over $\l$ which are again given by modified
Bessel functions, see \eqnref{2loopresult}.

%%%%%%%%%%%%%%%%%%%%%%%%%%%%%%%%%%%%%%%%%%%%%%%%%%
\section{\texorpdfstring{$n$}{n}-loop integrals}\label{app:nl-integrals}
%%%%%%%%%%%%%%%%%%%%%%%%%%%%%%%%%%%%%%%%%%%%%%%%%%
The calculation proceeds along the lines of the $2$-loop
integral discussed in Appendix B. The integral
(\ref{n-loop-int}) is given by $\Pi ^{n \text{ npl-ins.} }(p) = \l
^2 \sum\limits_{\eta = \pm 1} J_n (p)$ with
\begin{align}
J_n (p) & \equiv \intk
\frac{ \re ^{\ri \eta  k \p} }{ \left( \k ^2 \right)^n
\, \left[ k^2+m^2 + \frac{a^2}{k^2} \right] ^{n+1} }
\nonumber \\
&= \frac{1}{ 2^{n+1} \th^{2n}}
\sum\limits_{\zeta_1,\ldots,\zeta_{n+1}=\pm1} \intk \,
\frac{\re^{\ri\eta  k\p}}{\left( k^2\right)^n} \,
\prod\limits_{i=1}^{n+1} \left( \frac{ 1+\zeta_i\frac{m^2}{2M^2}
}{k^2+\frac{m^2}{2}+\zeta_iM^2} \right) \,.
\end{align}
We need a total of $n+2$ Schwinger parameters $\alpha_i$ to parameterize the denominators of the integrand~\cite{Leibbrandt:1994}:
\begin{align}
\inv{k^2+\frac{m^2}{2}+\zeta_iM^2}&=\int\limits_0^{\infty}d\alpha_i
\, \re^{-\alpha_i\left(k^2+\frac{m^2}{2}+\zeta_iM^2\right)}\,,
\quad\hspace{2.1cm} {\rm for} \ \, i \in \{ 1, \dots , n+1 \}
\,,\nonumber\\
\inv{(k^2)^n}&=\inv{\Gamma(n)}\int\limits_0^{\infty}d\alpha_{n+2}
\,  (\alpha_{n+2})^{n-1} \re^{ -\alpha_{n+2}k^2 }\,, \qquad {\rm for}
\ \, k^2 >0 \,.
\end{align}
We perform the change of variables $(\alpha_1,\ldots,\alpha_{n+2})\to(\xi_1,\ldots,\xi_{n+1},\l)$ with
\begin{align}\label{defalpha}
\begin{array}{ll}
\alpha_1=\l\prod\limits_{i=1}^{n+1}\xi_i \,, \quad
\alpha_2=\l(1-\xi_1)\prod\limits_{i=2}^{n+1}\xi_i \,, \quad  &
 \dots \, , \quad
\alpha_k=\l(1-\xi_{k-1})\prod\limits_{i=k}^{n+1}\xi_i\,,
\\
& \dots \, , \quad \alpha_{n+2}=\l(1-\xi_{n+1})\,,
\end{array}
\end{align}
where $\xi_i\in[0,1]$ and $\l\in[0,\infty [$. The integration
measure transforms as
\begin{align}
\prod\limits_{i=1}^{n+2}d\alpha_i &=\l^{n+1}
\prod\limits_{l=1}^{n}\left(\xi_{l+1}\right)^l \, d\l \,
\prod\limits_{j=1}^{n+1}d\xi_j\,,
\end{align}
and we have $\sum\limits_{i=1}^{n+2}\alpha_i=\l$. The integration over $k$ can be carried out by completing the square in the exponent so that we arrive at
\begin{align}
J_n&=\inv{\th^{2n}2^{n+1}(4\pi)^2\Gamma(n)}
\sum\limits_{\zeta_1,\ldots,\zeta_{n+1}}
\prod\limits_{i=1}^{n+1}\left(1+\zeta_i\tfrac{m^2}{2M^2}\right)
\int\limits_0^{\infty}d\l \, \l^{2n-2}
\prod\limits_{j=1}^{n+1}\int\limits_0^1d\xi_j \, \prod\limits_{l=1}^{n}\left(\xi_{l+1}\right)^l \nonumber\\
&\quad\; \qquad \times (1 - \xi_{n+1} )^{n-1}
\exp\left[-\frac{\p^{\, 2}}{4\l} -\l \xi_{n+1} \frac{m^2}{2} -
\left( \zeta_1 \alpha_1 + \ldots + \zeta_{n+1} \alpha_{n+1}
\right) M^2 \right],
\end{align}
where the coefficients $\alpha_1, \dots,\alpha_{n+1}$ in the exponent are functions of the variables $\l, \xi_1, \dots, \xi_{n+1}$ according to \eqnref{defalpha}.
The sum over the $\zeta_i$ can be expressed in terms of hyperbolic functions:
\begin{align}\label{jnp}
J_n&=\inv{\th^{2n}(4\pi)^2\Gamma(n)} \, \prod\limits_{j=1}^{n+1}
\int\limits_0^1 d\xi_j \, (1 - \xi_{n+1} )^{n-1} \,
\prod\limits_{l=1}^{n}\left(\xi_{l+1}\right)^l
\int\limits_0^{\infty}d\l \, \l^{2n-2}
\\
&\qquad \qquad \qquad \times
\re^{-\frac{\p^{\,2}}{4\l}-\l\xi_{n+1}\tfrac{m^2}{2}} \,
\prod\limits_{i=1}^{n+1} \left[\cosh\left( \alpha_i M^2 \right)
-\frac{m^2}{2M^2} \, \sinh\left( \alpha_i  M^2 \right)\right] \, .
\nonumber
\end{align}
Integration over $\xi_1,\ldots,\xi_{n+1}$ yields a sum of
integrals over $\l$ which are once more given by modified Bessel functions.

%%%%%%%%%%%%%%%%%%%%%%%%%%%%%%%%%%%%%%%%%%%%%%%%%%%%%%%%%%

%%%%%%%%%%%%%%%%%%%%%%%%%%%%%%%%%%%%%%%%%%%%%%%%%%%%%%%%%%%
% \bibliographystyle{custom1}
% \bibliography{articles,books,unpublished}

\begin{thebibliography}{10}
\expandafter\ifx\csname url\endcsname\relax
  \def\url#1{{\tt #1}}\fi
\expandafter\ifx\csname urlprefix\endcsname\relax\def\urlprefix{URL }\fi
\providecommand{\eprint}[2][]{\url{#2}}

\bibitem{Minwalla:1999}
S.~Minwalla, M.~Van~Raamsdonk and N.~Seiberg, {\it Noncommutative perturbative
  dynamics\/}, {\it JHEP\/} {\bf 02} (2000) 020,
  \href{http://www.arxiv.org/abs/hep-th/9912072}{{\tt [arXiv:hep-th/9912072]}}.
\\
% \bibitem{Raamsdonk:2000}
M.~Van~Raamsdonk and N.~Seiberg, {\it Comments on noncommutative perturbative
  dynamics\/}, {\it JHEP\/} {\bf 03} (2000) 035,
  \href{http://www.arxiv.org/abs/hep-th/0002186}{{\tt [arXiv:hep-th/0002186]}}.

\bibitem{Micu:2000}
A.~Micu and M.~M. Sheikh~Jabbari, {\it Noncommutative {$\Phi^4$} theory at two
  loops\/}, {\it JHEP\/} {\bf 01} (2001) 025,
  \href{http://www.arxiv.org/abs/hep-th/0008057}{{\tt [arXiv:hep-th/0008057]}}.

\bibitem{Rivasseau:2007a}
V.~Rivasseau, {\it Non-commutative renormalization\/}, in {\it
Quantum Spaces - Poincar\'{e} Seminar 2007}, B. Duplantier and V.
Rivasseau, eds. (Birkh\"auser Verlag, 2007),
% S\'{e}minaire Poincar\'{e} X}, {\it Prog. Math. Phys.} {\bf 53} (2007) 19,
  \href{http://www.arxiv.org/abs/0705.0705}{{\tt [arXiv:0705.0705]}}.

\bibitem{Douglas:2001}
M.~R. Douglas and N.~A. Nekrasov, {\it Noncommutative field theory\/}, {\it
  Rev. Mod. Phys.\/} {\bf 73} (2001) 977--1029,
  \href{http://www.arxiv.org/abs/hep-th/0106048}{{\tt [arXiv:hep-th/0106048]}}.
\\
% \bibitem{Szabo:2001}
R.~J. Szabo, {\it Quantum field theory on noncommutative spaces\/}, {\it Phys.
  Rept.\/} {\bf 378} (2003) 207--299,
  \href{http://www.arxiv.org/abs/hep-th/0109162}{{\tt [arXiv:hep-th/0109162]}}.

\bibitem{Nicolini:2008}
P.~Nicolini, {\it {Noncommutative Black Holes, The Final Appeal To Quantum
  Gravity: A Review}\/}, \href{http://www.arxiv.org/abs/0807.1939}{{\tt
  [arXiv:0807.1939]}}.

\bibitem{Rivasseau:2008a}
R.~Gurau, J.~Magnen, V.~Rivasseau and A.~Tanasa, {\it A translation-invariant
  renormalizable non-commutative scalar model\/},
  \href{http://www.arxiv.org/abs/0802.0791}{{\tt [arXiv:0802.0791]}}.

\bibitem{Tanasa:2008d}
A.~Tanasa, {\it {Scalar and gauge translation-invariant noncommutative models}\/},
  \href{http://www.arxiv.org/abs/0808.3703}{{\tt [arXiv:0808.3703]}}.

\bibitem{Grosse:2003}
H.~Grosse and R.~Wulkenhaar, {\it Renormalisation of $\phi^4$ theory on
  noncommutative $\mathbb{R}^2$ in the matrix base\/}, {\it JHEP\/} {\bf 12} (2003)
  019, \href{http://www.arxiv.org/abs/hep-th/0307017}{{\tt
  [arXiv:hep-th/0307017]}}.
\\
% \bibitem{Grosse:2004b}
H.~Grosse and R.~Wulkenhaar, {\it Renormalisation of $\phi^4$ theory on
  noncommutative $\mathbb{R}^4$ in the matrix base\/}, {\it Commun. Math. Phys.\/} {\bf
  256} (2005) 305--374, \href{http://www.arxiv.org/abs/hep-th/0401128}{{\tt
  [arXiv:hep-th/0401128]}}.

\bibitem{Blaschke:2007b}
D.~N. Blaschke, H.~Grosse and M.~Schweda, {\it Non-Commutative $U(1)$ Gauge
  Theory on $\mathbb{R}^4_\th$ with Oscillator Term and {BRST} Symmetry\/}, {\it Europhys.
  Lett.\/} {\bf 79} (2007) 61002,
  \href{http://www.arxiv.org/abs/0705.4205}{{\tt [arXiv:0705.4205]}}.

\bibitem{Grosse:2008a}
H.~Grosse and F.~Vignes-Tourneret, {\it Minimalist translation-invariant
  non-commutative scalar field theory\/},
  \href{http://www.arxiv.org/abs/0803.1035v1}{{\tt [arXiv:0803.1035v1]}}.

\bibitem{Blaschke:2008a}
D.~N. Blaschke, F.~Gieres, E.~Kronberger, M.~Schweda and M.~Wohlgenannt, {\it
  {Translation-invariant models for non-commutative gauge fields}\/}, {\it J.
  Phys.\/} {\bf A41} (2008) 252002,
  \href{http://www.arxiv.org/abs/0804.1914}{{\tt [arXiv:0804.1914]}}.

\bibitem{Rivasseau:2005a}
V.~Rivasseau, F.~Vignes-Tourneret and R.~Wulkenhaar, {\it Renormalization of
  noncommutative $\phi^4$-theory by multi-scale analysis\/}, {\it Commun. Math.
  Phys.\/} {\bf 262} (2006) 565--594,
  \href{http://www.arxiv.org/abs/hep-th/0501036}{{\tt [arXiv:hep-th/0501036]}}.
\\
% \bibitem{Rivasseau:2005b}
R.~Gurau, J.~Magnen, V.~Rivasseau and F.~Vignes-Tourneret, {\it Renormalization
  of non-commutative $\phi^4_4$ field theory in x space\/}, {\it Commun. Math.
  Phys.\/} {\bf 267} (2006) 515--542,
  \href{http://www.arxiv.org/abs/hep-th/0512271}{{\tt [arXiv:hep-th/0512271]}}.

\bibitem{workinprogress}
 Work in progress.

\bibitem{Maggiore:2005}
M.~Maggiore, {\it {A Modern Introduction to Quantum Field
Theory}\/}, New York:
  Oxford University Press (Oxford Series in Physics, 12.), 2005.

\bibitem{Rim:2003}
C.~Rim, Y.~Seo and J.~H. Yee, {\it {Perturbation theory of the space-time
  non-commutative real scalar field theories}\/}, {\it Phys. Rev.\/} {\bf D70}
  (2004) 025006, \href{http://www.arxiv.org/abs/hep-th/0312308}{{\tt
  [arXiv:hep-th/0312308]}}.

\bibitem{Tanasa:2008a}
J.~B. Geloun and A.~Tanasa, {\it {One-loop $\beta$ functions of a
  translation-invariant renormalizable noncommutative scalar model}\/},
  \href{http://www.arxiv.org/abs/0806.3886}{{\tt [arXiv:0806.3886]}}.

\bibitem{Tanasa:2008b}
A.~Tanasa, {\it {Parametric representation of a translation-invariant
  renormalizable noncommutative model}\/},
  \href{http://www.arxiv.org/abs/0807.2779}{{\tt [arXiv:0807.2779]}}.

\bibitem{Tanasa:2008c}
J.~Magnen, V.~Rivasseau and A.~Tanasa, {\it {Commutative limit of a
  renormalizable noncommutative model}\/},
  \href{http://www.arxiv.org/abs/0807.4093}{{\tt [arXiv:0807.4093]}}.

\bibitem{Gradshteyn:2007}
I.~S. Gradshteyn and I.~M. Ryzhik, {\it Table of Integrals, Series and
  Products\/}, Academic Press, seventh edition, 2007.

\bibitem{Leibbrandt:1994}
G.~Leibbrandt, {\it Noncovariant Gauges: Quantization of
{Yang-Mills} and
  {Chern-Simons} Theory in axial type Gauges\/}, Singapore: World Scientific,
  1994.

\end{thebibliography}
\end{document}